\documentclass[twocolumn,10pt]{article}
\usepackage[a4paper,margin=1.8cm,top=2.1cm,bottom=2.1cm,columnsep=0.7cm]{geometry}
\usepackage{lmodern}\usepackage[T1]{fontenc}
\usepackage{amsmath,amssymb,bm,mathtools}
\usepackage{graphicx}
\usepackage{xcolor}
\usepackage[font=small,labelfont=bf,labelsep=period]{caption}
\usepackage{subcaption}
\usepackage[numbers,sort&compress]{natbib}
\usepackage{fancyhdr}
\usepackage{titlesec}
\usepackage{enumitem}
\usepackage[colorlinks=true,linkcolor=blue!50!black,citecolor=blue!50!black,urlcolor=blue!50!black]{hyperref}

\titleformat{\section}{\normalfont\sffamily\bfseries\large}{}{0pt}{}
\titleformat{\subsection}{\normalfont\sffamily\bfseries}{}{0pt}{}
\titleformat{\paragraph}[runin]{\normalfont\bfseries}{}{0pt}{}[.]
\titlespacing*{\section}{0pt}{1.2em}{0.5em}
\titlespacing*{\subsection}{0pt}{1em}{0.3em}
\titlespacing*{\paragraph}{0pt}{0.6em}{0.6em}

\newcommand{\Meff}{M_{\mathrm{eff}}}
\newcommand{\Muu}{M_{uu}}
\newcommand{\Mww}{M_{ww}}
\newcommand{\Muw}{M_{uw}}
\newcommand{\Mwu}{M_{wu}}
\newcommand{\In}{\mathrm{In}}
\newcommand{\T}{^{\mathsf{T}}}
\newcommand{\KK}{\mathcal{K}}

\title{\vspace{-1.2em}\sffamily\bfseries\LARGE Interactions with hidden traits induce diversification and conceal its origin}
\author{\normalsize Vikas Kumar Jha$^{\dagger}$, Praveen V$^{\dagger}$, Suman Chakraborty, and Vaibhav Madhok$^{\ast}$\\[0.4em]
\small\itshape Department of Physics, Indian Institute of Technology Madras, Chennai, India 600036\\[0.3em]
\small\upshape $^{\dagger}$These authors contributed equally.\quad $^{\ast}$Corresponding author: \texttt{madhok@physics.iitm.ac.in}}
\date{}

\begin{document}
\twocolumn[
\begin{@twocolumnfalse}
\maketitle
\vspace{-2em}
\noindent\rule{\textwidth}{0.4pt}
\small
\noindent\textbf{Abstract.}
Adaptive diversification is often inferred from a small number of measurable traits, even though evolutionary dynamics arise from many interacting phenotypic dimensions. Unmeasured traits may therefore do more than bias inference: through their interactions, they may determine whether diversification occurs at all. Here we develop a theory of the onset and attribution of adaptive diversification by partitioning phenotype space into focal and hidden sectors. Eliminating the hidden sector yields an exact effective operator. Coupling to hidden traits that are stabilizing in isolation can destabilize an otherwise stable phenotype and generate a diversifying composite mode: the interaction creates diversification rather than merely strengthening it. Conversely, when the hidden sector is itself diversifying, the focal traits diversify as a correlated response to a hidden-driven process: no effective description in the focal traits, not even the exact one obtained by eliminating the hidden sector, predicts the variation they display, and stabilizing the focal traits reduces it without removing it. These effects become strongest near the hidden sector's own threshold, so a weakly stabilized hidden trait can have a large effect even with modest coupling to the focal traits, and they can generate correlations between focal traits that have no direct interaction. Onset and observability are distinct: systems with the same leading eigenvalue can project very differently onto the measured traits, so the theory tracks both the full operator and the focal-hidden partition. Individual-based simulations recover the predicted threshold and variance law, coupling-induced trait splitting, and the reversal of induced correlations. Interactions with unmeasured traits can both create adaptive diversification and determine whether the diversification seen in measured phenotypes can be attributed to them.

\medskip
\noindent\textbf{One-sentence summary.} Interactions with hidden phenotypic traits can induce diversification in otherwise stable populations, amplify it near criticality, and drive diversification that measured traits display but cannot explain. 
\par\noindent\rule{\textwidth}{0.4pt}
\vspace{1em}
\end{@twocolumnfalse}
]

\section{Introduction}

Phenotypes or traits are the physical, physiological, or behavioral expressions of an organism that arise from the interaction between its genetic makeup (genotype) and its environment. Complex interactions among tens of thousands of genes within the cells of an organism can give rise to a vast number of anatomical, physiological, and behavioral traits. Among these large number of traits, only a few may be readily identifiable or directly measurable, such as height, body size, and coloration. We refer to these as \emph{observable traits}. In contrast, many traits are not directly observed in a given study and manifest only through internal or physiological characteristics---metabolic rate, hormone and gene-expression levels, immune parameters, or microbiome composition. We refer to these as \emph{hidden traits}. Interestingly, these traits are not necessarily independent of one another; rather, they can influence each other and interact as components of a complex system, potentially giving rise to rich emergent behavior \citep{pigliucci2003,hallgrimsson2019,chakraborty2023}.

According to Darwin's theory of evolution, the enormous diversity of phenotypes observed in nature has emerged through the gradual modification and diversification of ancestral populations through processes such as mutation and natural selection \citep{darwin1859,barrett2008,sniegowski1995}. Mutations generate variation in phenotypic characteristics, while natural selection determines which of these variations confer advantages for survival and reproduction under a given environmental condition. Over many generations, the accumulation of such variations and their differential survival and reproduction can lead to the divergence of populations and, ultimately, the formation of new species \citep{darwin1859,coyne2004,nei2011}.

Surprisingly, natural populations often exhibit far more phenotypic variation than might be expected from a simple mutation--selection picture, and evolutionary lineages can repeatedly undergo diversification \citep{geritz1998}. Theoretical studies have shown that frequency-dependent competition favors rare phenotypes and can maintain variation or generate diversification, but one-dimensional models often require strong niche differentiation \citep{dieckmann1999,doebeli2010}. In a multidimensional phenotype, weak interactions can combine into a direction along which competitive curvature exceeds stabilizing curvature. This mechanism makes diversification increasingly likely as trait dimension grows and has been demonstrated analytically and in individual-based models \citep{doebeli2010,debarre2014,ispolatov2016,doebeli2017}.

However, to the best of our knowledge, theoretical investigations of evolutionary diversification have focused primarily on observable or explicitly modeled traits. The role of a large number of unobserved traits, and in particular the effects of their complex interactions with observable traits, on diversification remains largely unexplored. That unmeasured traits correlated with the measured ones can distort estimates of selection is a classic concern in quantitative genetics \citep{lande1983,mitchellolds1987}. What has not been examined is their effect on the onset of diversification under frequency-dependent competition. In this paper, we address this gap by asking how unmeasured traits alter the diversification inferred from measured traits.

The answer follows from a block elimination of the competition-minus-stabilization operator. Integrating out the hidden block produces an exact focal self-energy whose sign is determined by the stability of the hidden sector. Stable hidden modes amplify focal diversification and can be eliminated exactly into an effective focal description; hidden modes that are themselves unstable toward diversification cannot: they drive variation in the focal traits that no focal-only description predicts. Near marginal stability of the hidden sector, this effect becomes highly sensitive. We test these predictions using individual-based dynamics and distinguish robust sign-level predictions from quantitative effect sizes that remain model-dependent. Moreover, the underlying criterion for diversification can be formulated in terms of a spectral-edge crossing. Closely related spectral-edge crossings organize the stability of large ecological communities \citep{may1972,allesina2012}, appear in Roy's largest-root statistic for competing quadratic forms \citep{roy1953,johnstone2008}, characterize spiked random-matrix transitions \citep{baik2005,benaych2011,johnstone2017,wang2017}, and are associated with topological transitions in confined random landscapes \citep{fyodorov2004,fyodorov2012,auffinger2013,ros2019,fyodorov2016}. These connections provide a mature mathematical framework for studying finite-size fluctuations and low-rank perturbations. A useful analogy is provided by block elimination in chaos-assisted tunneling \citep{tomsovic1994}. In that setting, eliminating a large chaotic sector dresses the effective coupling between regular states. Here, eliminating the hidden traits dresses the effective diversification operator in the focal trait space. The analogy is structural: in both cases, eliminating a sector generates an effective operator through a Schur complement. Block elimination of this kind is also used in community ecology, where eliminating one group of species yields an effective competition matrix for the other \citep{ferrera2017}; here the eliminated block consists of unmeasured traits, and the question is what an observer of the remaining traits can infer.

\begin{figure*}[t]
\centering
\includegraphics[width=\textwidth]{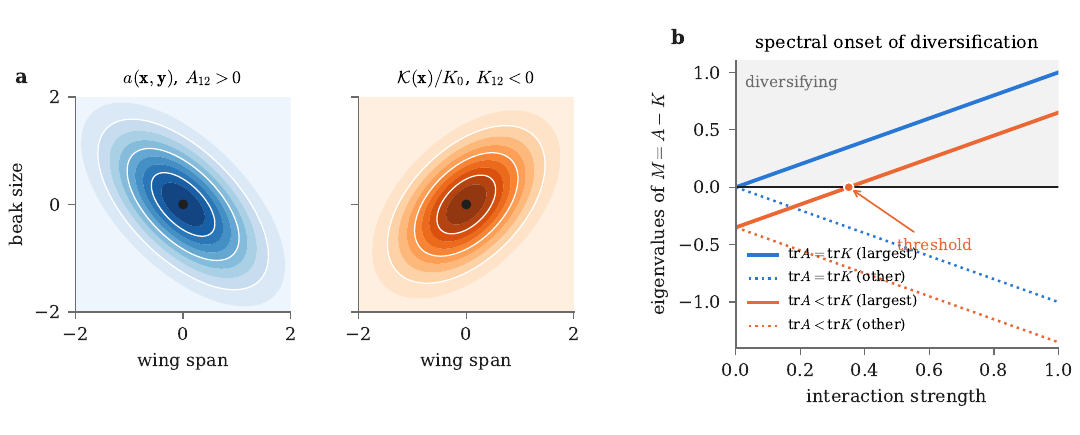}
\caption{\textbf{Schematic representation of geometry and spectral onset of diversification:} (a) An off-diagonal entry of $A$ ($A_{12}>0$ shown) rotates the contours of ecological similarity around each individual, whereas an off-diagonal entry of $K$ ($K_{12}<0$ shown) rotates the contours of the fixed environmental optimum. The mean value of each trait is taken to be zero and scaled to match the axes. (b) Interactions split the eigenvalues of $M$ while preserving their sum (drawn for $M=\begin{psmallmatrix}-\varepsilon&s\\s&-\varepsilon\end{psmallmatrix}$, eigenvalues $-\varepsilon\pm s$). For balanced total curvature, i.e., $\operatorname{tr}A=\operatorname{tr}K$, any nonzero splitting generates a positive eigenvalue, whereas stronger stabilization, i.e., $\operatorname{tr}A<\operatorname{tr}K$, requires the interaction strength to exceed a finite threshold \citep{doebeli2010}.}
\label{fig:schematic}
\end{figure*}

\section{Logistic Growth Model and Spectral Criterion for Diversification}

A well-established mathematical framework for describing evolutionary dynamics is the logistic growth model with phenotype-dependent competition and carrying capacity \citep{verhulst1838,roughgarden1976,macarthur1967}. According to the logistic growth model, the population density $\phi$ evolves according to
\begin{equation}
\partial_t\phi(\bm x,t)=r\,\phi(\bm x,t)\left[1-\frac{\int a(\bm x,\bm y)\,\phi(\bm y,t)\,d\bm y}{\KK(\bm x)}\right].
\label{eq:logistic}
\end{equation}
In this framework, the population dynamics are governed by two competing effects: the competition kernel $a(\bm x,\bm y)$, which quantifies the strength of competition between individuals with phenotypes $\bm x$ and $\bm y$, and the carrying-capacity function $\KK(\bm x)$, which describes the environmental suitability for individuals with phenotype $\bm x$.

We use the multidimensional Gaussian competition model \citep{doebeli2010} to sketch the basic idea, with competition kernel and carrying capacity given by
\begin{equation}
\begin{aligned}
a(\bm x,\bm y)&=\exp\!\left[-(\bm x-\bm y)\T A(\bm x-\bm y)\right],\\
\KK(\bm x)&=K_0\exp\!\left(-\bm x\T K\bm x\right),
\end{aligned}
\label{eq:gauss}
\end{equation}
where $A$ and $K$ are symmetric positive-definite matrices and $\bm x$ is a column vector. The off-diagonal entries of $A$ and $K$ have distinct biological interpretations. The off-diagonal elements of $A$ influence the orientation of the axes of ecological similarity in phenotype space, whereas those of $K$ describe correlations among phenotypic directions in the stabilizing selection imposed by the environment (Fig.~\ref{fig:schematic}a). Positive definiteness ensures that competition is strongest between identical phenotypes and that the carrying capacity has a unique maximum at $\bm x=0$. The competition kernel therefore introduces frequency-dependent competition: individuals with similar phenotypes compete more strongly with one another, so that common phenotypes can experience stronger competitive pressure than rare phenotypes. This frequency dependence can generate disruptive selection and thereby promote diversification. In contrast, the carrying-capacity function represents stabilizing selection, favoring phenotypes closer to the environmental optimum $\bm x=0$. When the disruptive force of competition exceeds the stabilizing effect of the carrying capacity along some direction $\bm q$, i.e.\
\begin{equation}
\bm q\T A\bm q>\bm q\T K\bm q,
\label{eq:qcond}
\end{equation}
variation is maintained along $\bm q$; in the Gaussian model this condition is both necessary and sufficient \citep{doebeli2010}.

We define the competition-minus-stabilization form\footnote{Under a change of trait coordinates $\bm x\mapsto S\bm x$, $M$ transforms by congruence, $M\mapsto S\T MS$, as do its Schur complements below under block-diagonal $S$. Every statement in this paper concerns the inertia of such matrices, which is congruence-invariant; we nevertheless refer to $M$ and $\Meff$ as operators where convenient.}
\begin{equation}
M\equiv A-K .
\label{eq:M}
\end{equation}
Diversification sets in when there exists a direction in trait space along which competition exceeds stabilization, i.e.\ when $M$ fails to be negative semi-definite:
\begin{equation}
\lambda_{\max}(M)>0,
\label{eq:criterion}
\end{equation}
where $\lambda_{\max}$ denotes the largest eigenvalue of $M$. Equivalently $n_+(M)\ge1$, with $n_+$ the positive index of inertia, and the number of diversifying directions is
\begin{equation}
N_{\rm div}=n_+(M).
\label{eq:ndiv}
\end{equation}
By Sylvester's law of inertia \citep{horn1985} the signature of $M$ is invariant under congruence, so Eqs.~\eqref{eq:criterion}--\eqref{eq:ndiv} are independent of the choice of trait coordinates, even though the eigenvalues of $M$ themselves are not. The diversifying directions are the generalized eigenvectors $A\bm v=d\,K\bm v$ with $d>1$, along which the dynamics decouples into independent one-dimensional models with $a/k=d$ \citep{doebeli2010}. Diversification begins when the largest eigenvalue of $M$ becomes positive (Fig.~\ref{fig:schematic}b). Adding traits changes how likely this crossing is and how abruptly it happens, but the test remains the sign of that eigenvalue.

If the ecological properties of an organism are determined by multiple interacting traits, the conditions under which frequency-dependent selection generates diversity are relaxed to the point where they are easily satisfied in high-dimensional phenotype spaces \citep{doebeli2010}: off-diagonal interactions in either $A$ or $K$ carry populations on the brink of diversification across the threshold (Fig.~\ref{fig:schematic}b).

\section{Diversification and Stabilization of Observable Traits by Hidden Traits}

Thus far, the phenotype space has been treated as a general trait space, without distinguishing between observable and hidden traits. We now partition the phenotype $\bm x$ into two components: observable (focal) traits, denoted by coordinates $\bm u$, and hidden traits, denoted by coordinates $\bm w$. Accordingly, the competition-minus-stabilization operator can be written in block form as
\begin{equation}
M=\begin{pmatrix}\Muu&\Muw\\ \Mwu&\Mww\end{pmatrix}.
\label{eq:block}
\end{equation}
When the hidden block $\Mww$ is nonsingular, block elimination gives the exact effective operator in the observable trait space,
\begin{equation}
\Meff\equiv M/\Mww=\Muu-\Muw\Mww^{-1}\Mwu=\Muu+\Sigma,
\label{eq:schur}
\end{equation}
where $\Sigma\equiv\Muw(-\Mww)^{-1}\Mwu$ is the correction term induced by the hidden sector. It can be interpreted as a hidden-sector resolvent dressed by the couplings between the observable and hidden traits. Note that the reduction must be performed on the full competition-minus-stabilization operator $M=A-K$. In general, eliminating the hidden blocks of $A$ and $K$ separately and then subtracting does not reproduce the effective operator obtained by eliminating the hidden block of $M$.

\begin{figure*}[t]
\centering
\includegraphics[width=\textwidth]{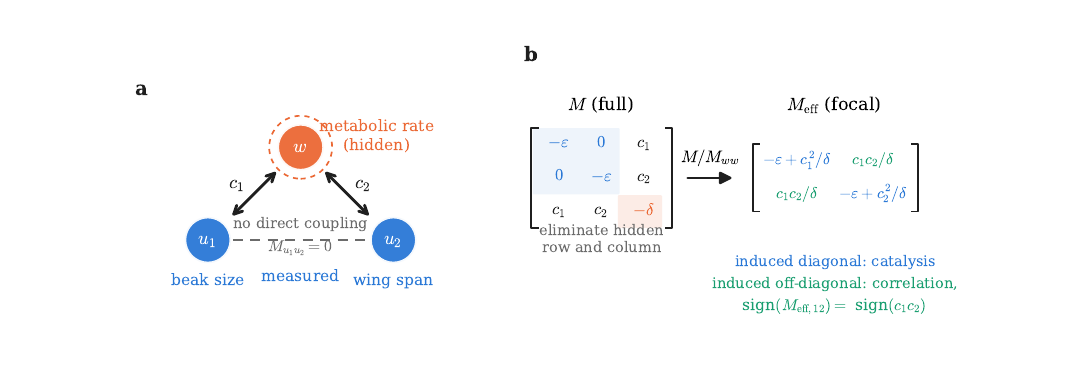}
\caption{\textbf{Schematic to illustrate the attribution mechanism:} (a) The measured traits, beak size and wing span, have no direct coupling but both interact with the hidden trait $h$ (e.g., metabolic rate). (b) For hidden margin $-\delta$ and couplings $c_1,c_2$, eliminating the hidden trait adds $c_i^2/\delta$ to the focal diagonal entries and $c_1c_2/\delta$ to the focal off-diagonal entry.}
\label{fig:mechanism}
\end{figure*}

Eq.~\eqref{eq:schur} provides a direct mechanism by which hidden traits influence the dynamics of observable traits. Two observable traits may have no direct interaction, yet become effectively coupled if both interact with a common hidden trait (Fig.~\ref{fig:mechanism}). The induced diagonal terms modify the effective diversification tendency of individual observable traits, while the induced off-diagonal terms generate effective associations between observable traits. The exact relation between diversification in the full and reduced systems follows from the Haynsworth inertia additivity formula \citep{haynsworth1968}:
\begin{equation}
\In(M)=\In(\Mww)+\In(\Meff),
\label{eq:haynsworth}
\end{equation}
where $\In$ denotes the inertia, i.e., the numbers of positive, negative, and zero eigenvalues. This relation leads to three useful predictions concerning diversification and its attribution.

\paragraph{Stable hidden sector}
If $\Mww\prec0$\footnote{Throughout this work, $T\prec0$ and $T\succ0$ denote that $T$ is negative definite and positive definite, respectively.}, then $(-\Mww)^{-1}\succ0$ and therefore $\Sigma\succeq0$. Thus, hidden directions that are stable in isolation can only increase the effective focal operator in the Loewner order. Moreover,
\[
n_+(M)=n_+(\Meff),
\]
so every positive direction of the full system is represented in the reduced observable trait space. Interestingly, as explained in Appendix~\ref{app:count}, when the focal sector is itself stable the number of induced diversifying directions is bounded by the rank of the focal--hidden coupling, and hence by the number of hidden traits involved.

\begin{figure}[t]
\centering
\includegraphics[width=\linewidth]{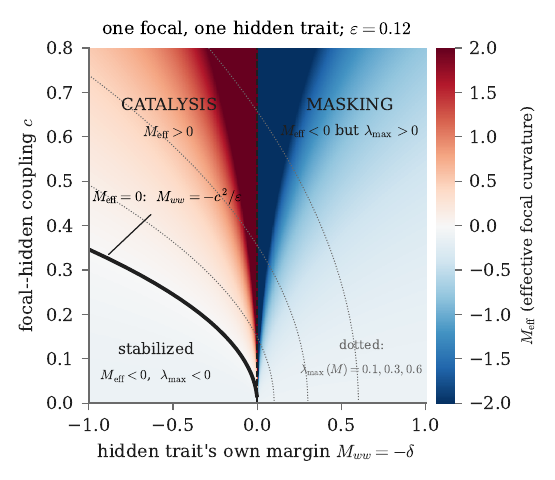}
\caption{\textbf{Catalysis vs.\ masking across the hidden-trait threshold:} For $\varepsilon=0.12$, the figure shows the two regimes for one focal trait coupled to one hidden trait. The dashed vertical line marks the hidden-trait threshold $\Mww=0$; the solid black curve is $\Meff=0$, i.e.\ $\Mww=-c^2/\varepsilon$. For $\Mww<0$ (hidden trait stable on its own) the focal trait is stabilized below the curve and catalyzed above it. For $\Mww>0$ (hidden trait diversifying on its own) $\Meff<0$ throughout, yet the full system diversifies for every $c$: this is the masking regime, in which the focal trait displays variation that no focal description predicts. Below the curve on the left the focal trait is stabilized ($\Meff<0$, $\lambda_{\max}<0$). Dotted contours: $\lambda_{\max}(M)$. In both regimes $|\Meff|$ grows as $\Mww\to0$.}
\label{fig:catmask}
\end{figure}

\paragraph{Diversifying hidden sector}
If $\Mww\succ0$, then $\Sigma\preceq0$ and hence
\[
n_-(M)=n_-(\Meff).
\]
In this case the positive directions of $M$ are inherited from the hidden block: $n_+(M)=n_+(\Mww)+n_+(\Meff)$ with $n_+(\Meff)\le n_+(\Muu)$, so a focal sector that is stable on its own contributes none. The focal traits nevertheless diversify, because every diversifying direction has a nonzero focal component whenever $\Muw\neq0$: from $(M-\lambda_+)\bm v=0$, $\bm v_u=(\lambda_+I-\Muu)^{-1}\Muw\bm v_w$. An observer measuring only $\bm u$ therefore sees variance; what no description confined to $\bm u$ supplies is its cause, since the naive focal form $\Muu$ and the exact effective form $\Meff$ are both negative definite. We call this regime \emph{masking}: the focal traits display diversification that is driven elsewhere. This is consistent with the observation of \citet{doebeli2010} that diversification along a composite direction appears as positive variance in generally many of the original trait components. The contrast with catalysis is operational rather than visual. In both regimes the focal variance is positive, but in catalysis it is extinguished by stabilizing the focal trait (for one focal and one hidden trait, at $\varepsilon=c^2/\delta$ in the notation of Eq.~\eqref{eq:toy} below), whereas in masking it persists for every focal margin, because $\det M=\varepsilon\delta-c^2<0$ for all $\varepsilon>0$ when $\delta<0$ (Fig.~\ref{fig:catmask}). Equivalently, elimination is informative when the eliminated block is the stable one: when $\Mww\succ0$ and $\Muu\prec0$ the natural effective description is obtained by eliminating the focal block instead, $M/\Muu=\Mww-\Mwu\Muu^{-1}\Muw\succeq\Mww$, and it lives in the hidden coordinates.

\paragraph{Mixed hidden sector}
If $\bm q_\alpha$ and $\mu_\alpha$ are the eigenvectors and eigenvalues of $\Mww$, respectively, then $\Mww\bm q_\alpha=\mu_\alpha\bm q_\alpha$ and $\bm b_\alpha=\Muw\bm q_\alpha$. Therefore, the correction term can be written as
\begin{equation}
\Sigma=-\sum_\alpha\frac{\bm b_\alpha\bm b_\alpha\T}{\mu_\alpha}.
\label{eq:modes}
\end{equation}
Each stable hidden mode ($\mu_\alpha<0$) therefore contributes a positive semidefinite term to $\Sigma$, whereas each diversifying hidden mode ($\mu_\alpha>0$) contributes a negative semidefinite term. The two definite-sector sign laws described above are thus recovered as limiting cases of this mode-resolved decomposition.

To illustrate our proposal, let us consider a simple setting with two stabilizing focal traits with no direct coupling and one stabilizing hidden trait that is coupled to both focal traits. Mathematically, the corresponding structure of $M$ is given by
\begin{equation}
\Muu=-\varepsilon I_2,\qquad \Muw=\begin{pmatrix}c_1\\c_2\end{pmatrix},\qquad \Mww=-\delta .
\label{eq:toy}
\end{equation}
Then
\begin{equation}
\Meff=-\varepsilon I_2+\frac1\delta\begin{pmatrix}c_1^2&c_1c_2\\c_1c_2&c_2^2\end{pmatrix}.
\label{eq:toyeff}
\end{equation}
The composite diversifying direction is
\begin{equation}
\hat{\bm c}=\begin{pmatrix}c_1\\c_2\end{pmatrix},
\label{eq:chat}
\end{equation}
and the eigenvalue along this direction is
\begin{equation}
\frac{c_1^2+c_2^2}{\delta}-\varepsilon .
\label{eq:toylambda}
\end{equation}
Therefore, diversification can happen along $\hat{\bm c}$ only when $c_1^2+c_2^2>\varepsilon\delta$. In contrast, in the direction orthogonal to $\hat{\bm c}$, the system remains stabilizing, with eigenvalue $-\varepsilon$. Note that here we have considered a stabilizing hidden trait, corresponding to $\delta>0$ and hence $\Mww=-\delta<0$. The coupling to this stable hidden trait can nevertheless induce diversification in the observable trait space whenever $c_1^2+c_2^2>\varepsilon\delta$. In contrast, if the hidden block is diversifying, $\delta<0$, the induced correction is negative semidefinite and the effective focal operator cannot acquire a positive eigenvalue. The full system still has exactly one diversifying direction, inherited from the hidden trait, and its focal component is again $\propto\hat{\bm c}$: the focal traits diversify along $\hat{\bm c}$ as a correlated response that no focal description predicts. This is the masking regime of Fig.~\ref{fig:catmask}. In either case a single hidden trait contributes at most one direction (Appendix~\ref{app:count}), while the direction orthogonal to $\hat{\bm c}$ remains stabilizing.

Interestingly, since it is the inverse of $\Mww$ that appears in the correction term $\Sigma$, near singular values of $\Mww$ the effect of focal--hidden coupling becomes more prominent. In our illustrative example the magnitude of the correction term
\begin{equation}
\Sigma=\frac1\delta\begin{pmatrix}c_1^2&c_1c_2\\c_1c_2&c_2^2\end{pmatrix}
\label{eq:toysigma}
\end{equation}
increases as the hidden mode approaches marginality, which is also evident from Fig.~\ref{fig:catmask}. The scenario when the hidden block $\Mww$ is singular is discussed in Appendix \ref{app:pole}.

The sign of $c_1c_2$ determines the orientation of the induced coupling and hence the direction of diversification. For $c_1=c_2=c$, the eigenvalue along the $(1,1)$ direction is $2c^2/\delta-\varepsilon$, which becomes positive when $2c^2>\varepsilon\delta$. Reversing the sign of one coupling changes the diversifying direction to $(1,-1)$, as also shown in Fig.~\ref{fig:sims}c. This sign reversal is particularly informative because it changes the orientation of the diversifying mode rather than merely modifying the magnitude of the marginal variance; such a change cannot be generated by a purely diagonal correction. We now verify these conclusions using individual-based simulation.

\section{Individual-based tests}

Here we test the static predictions against continuous-time birth--death simulations of Eq.~\eqref{eq:logistic} with Gaussian mutation. These dynamics include finite-population drift, mutation-induced variance, and nonlinear ecological feedback absent from the local quadratic reduction. The individual-based simulations are implemented as a stochastic birth--death process corresponding to the logistic model \citep{madhok2020}. Each individual $i$, characterized by its trait vector $\bm x_i$, is assigned a constant birth rate $\rho_i=1$, while its death rate is given by $\gamma_i=N_{\rm eff}(i)/\KK(\bm x_i)$. For a population of $N$ individuals, the effective density experienced by individual $i$ is $N_{\rm eff}(i)=\sum_{j\neq i}a(\bm x_j,\bm x_i)$, where $a(\bm x_j,\bm x_i)$ is the Gaussian competition kernel. The total event rate is therefore $U=\sum_i(\rho_i+\gamma_i)$. At each computational step, a birth or death event is selected stochastically with probability proportional to its corresponding rate. In a birth event, an offspring is added with a trait vector obtained by perturbing the parental trait vector by a Gaussian random displacement with small variance, whereas in a death event the selected individual is removed from the population. The simulation time is advanced according to the waiting time associated with the total rate $U$. Note that although our analytical treatment does not explicitly include mutation, the qualitative conclusions remain unchanged when mutation is introduced, as discussed by \citet{doebeli2010}. Unless stated otherwise, $K_0=500$ (so that $N\approx500$--$900$), the mutation step has standard deviation $\sigma_{\rm mut}=0.015$ per trait, runs last $2$--$6\times10^6$ events, reported variances are averaged over the second half of each run, and error bars are the spread over independent seeds. Simulations initialized with phenotypic spreads of $0.05$, $0.25$, and $0.50$ converged to comparable final variances, providing a basic equilibration check.

\begin{figure*}[t]
\centering
\includegraphics[width=\textwidth]{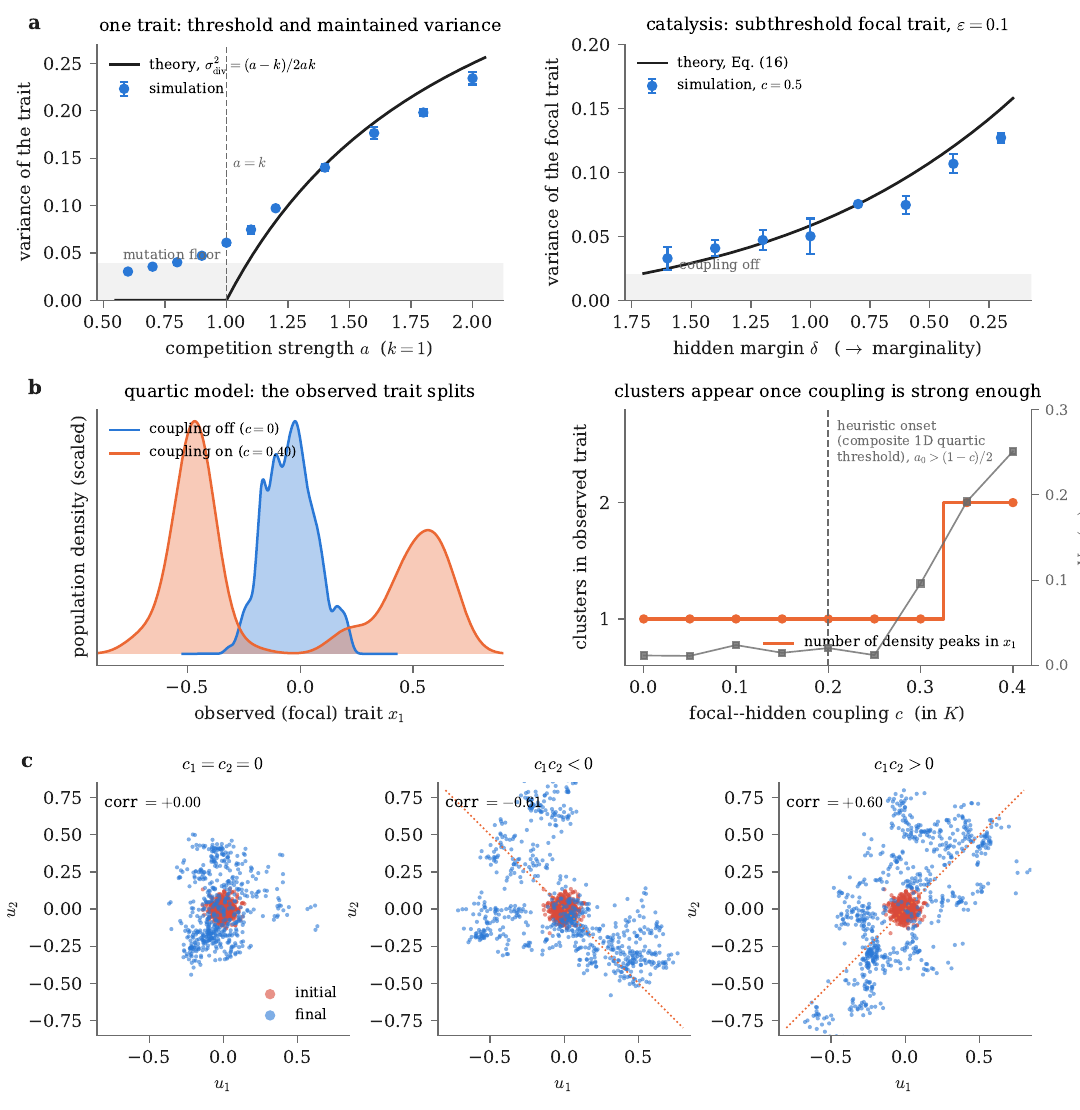}
\caption{\textbf{Simulation to test spectral and attribution predictions:} (a) Left: in one dimension the variance stays at the mutation floor below the threshold $a=k=1$ and follows Eq.~\eqref{eq:var1d} above it to within $\sim6\%$ ($\sigma_{\rm mut}=0.03$, $4\times10^6$ events, three seeds). Right: a subthreshold focal trait ($\varepsilon=0.1$) coupled to a stable hidden trait ($c=0.5$) acquires variance that grows as the hidden margin $\delta$ approaches zero and follows Eq.~\eqref{eq:var2d}; shaded band, variance with the coupling switched off. (b) Quartic model, $a(\bm x,\bm y)=\exp[-((\bm x-\bm y)\T A(\bm x-\bm y))^2]$, $\KK=K_0\exp[-(\bm x\T K\bm x)^2]$, with $A=a_0I$, $a_0=0.4$, and $K=\begin{psmallmatrix}1&-c\\-c&1\end{psmallmatrix}$: each trait is below the one-dimensional quartic threshold on its own. Left: kernel density of the observed trait $x_1$ with the coupling off and on. Right: number of density peaks in $x_1$ (orange) and $\operatorname{Var}(x_1)$ (grey, right axis) against $c$; the dashed line is the heuristic onset obtained by restricting the quartic model to the composite direction $(1,1)/\sqrt2$ and applying the one-dimensional quartic threshold of \citet{doebeli2010}, $a_0>\tfrac12(1-c)$. The measured onset lies at larger $c$, as expected for a criterion that is not exact beyond the quadratic model. (c) Two focal traits coupled to one hidden trait as in Eq.~\eqref{eq:toy}, $\varepsilon=0.1$, $\delta=0.2$, $|c_1|=|c_2|=0.4$, $\sigma_{\rm mut}=0.03$: near-zero correlation without coupling, positive correlation for $c_1c_2>0$, negative for $c_1c_2<0$ (dotted lines, predicted directions). Red, initial population; blue, after $3\times10^6$ events. Elimination shifts the eigenvalue of $\Meff$ along $(1,1)$ from $-0.1$ to $+1.5$; reversing one coupling selects $(1,-1)$.}
\label{fig:sims}
\end{figure*}

\paragraph{Variance for a single trait}
In one dimension, the Gaussian equilibrium above the diversification threshold has variance
\begin{equation}
\sigma^2_{\rm div}=\frac{a-k}{2ak}.
\label{eq:var1d}
\end{equation}
Below the threshold, finite mutation maintains a small but nonzero variance. The simulations reproduce both the onset of diversification at $a=k$ and the predicted variance scale above the threshold (Fig.~\ref{fig:sims}a).

\paragraph{Hidden-trait catalysis}
A subthreshold focal trait coupled to a stable hidden trait develops variance above the mutation floor, with the effect becoming stronger as the hidden trait approaches marginality, i.e., as $\delta$ decreases toward zero. For one focal trait $x$ and one hidden trait with $A=I$ and $M$ as in Eq.~\eqref{eq:toy2} below, the stationary population is Gaussian along the diversifying eigenvector of $M$ and concentrated across it (Appendix~\ref{app:gauss}), so that the focal variance is
\begin{equation}
\begin{aligned}
\operatorname{Var}(x)&=\frac12\,\frac{\lambda_+}{1-\lambda_+}\,\frac{c^2}{c^2+(\varepsilon+\lambda_+)^2},\\
\lambda_+&=\tfrac12\Big[-(\varepsilon+\delta)+\sqrt{(\varepsilon-\delta)^2+4c^2}\,\Big],
\end{aligned}
\label{eq:var2d}
\end{equation}
for $c^2>\varepsilon\delta$ and zero otherwise. The simulated focal variance follows Eq.~\eqref{eq:var2d} quantitatively across the hidden margin, down to the mutation floor at large $\delta$ (Fig.~\ref{fig:sims}a, right).

In the quartic model, in which the exponents in Eq.~\eqref{eq:gauss} are replaced by their squares, $a=\exp[-((\bm x-\bm y)\T A(\bm x-\bm y))^2]$ and $\KK=K_0\exp[-(\bm x\T K\bm x)^2]$ \citep{doebeli2010}, diversification produces distinct clusters rather than a broadened Gaussian distribution. The quadratic theory does not apply directly to this model---the curvatures at the optimum vanish, and the one-dimensional threshold lies at $a\approx0.25k$ rather than $a=k$ \citep{doebeli2010}---so we use it only as a qualitative test of the predicted direction and sign of the coupling effect: coupling two individually subthreshold traits changes the observed distribution from a single mode to two distinct modes, providing a clearer qualitative signature of diversification (Fig.~\ref{fig:sims}b).

\paragraph{Off-diagonal sign test}
For the construction in Eq.~\eqref{eq:toyeff}, the focal traits are uncoupled by design. When the hidden coupling is removed, their correlation remains close to zero. Couplings with $c_1c_2>0$ produce a positive correlation of approximately $+0.6$, whereas reversing the sign of one coupling produces a negative correlation of approximately $-0.6$. The sign reversal is consistent with the theoretical relation
\begin{equation}
\operatorname{sign}(\Sigma_{12})=\operatorname{sign}(c_1c_2),
\label{eq:signrule}
\end{equation}
see Fig.~\ref{fig:sims}c.

\begin{figure*}[t]
\centering
\includegraphics[width=\textwidth]{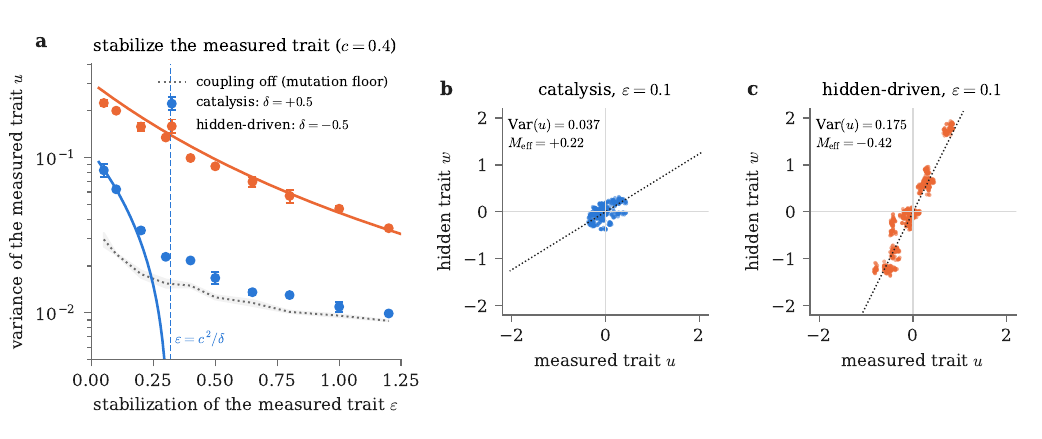}
\caption{\textbf{Catalysis and masking are separated by stabilizing the measured trait, not by looking at it:} one measured trait $u$ coupled ($c=0.4$) to one hidden trait that is either stable on its own ($\delta=+0.5$, blue) or diversifying on its own ($\delta=-0.5$, orange); $A=I$, $6\times10^6$ events, two seeds. (a) Variance of $u$ against its own stabilization $\varepsilon$. Catalyzed variance falls toward the mutation floor (grey band, coupling off) beyond $\varepsilon=c^2/\delta=0.32$ and reaches it by $\varepsilon\approx0.65$; hidden-driven variance decays as the focal share of the hidden-driven mode but never reaches the floor, while $\operatorname{Var}(w)$ stays between about $0.55$ and $0.8$ (not shown). Solid lines: Eq.~\eqref{eq:var2d}; dotted line with grey band: simulation with the coupling switched off.
 (b,c) Snapshots at $\varepsilon=0.1$ in the two regimes: the population is a line along the leading eigenvector of $M$ (dotted) with a visible spread in $u$ in both; $\Meff$ has opposite signs.}
\label{fig:masking}
\end{figure*}

\paragraph{Masking test}
One focal trait coupled to a hidden trait with $\delta<0$ retains variance above the mutation floor for every focal margin $\varepsilon$, whereas for $\delta>0$ the variance falls back toward the floor once $\varepsilon>c^2/\delta$ (Fig.~\ref{fig:masking}a; $c=0.4$, $|\delta|=0.5$, threshold $\varepsilon=0.32$). In both regimes the stationary population is a line in the $(u,w)$ plane with a visible shadow on the measured axis (Fig.~\ref{fig:masking}b,c), so the two are indistinguishable in a snapshot and distinguished by the $\varepsilon$-sweep. The hidden-driven focal variance follows Eq.~\eqref{eq:var2d} with $\delta<0$, i.e.\ it decays as the focal share $c^2/(c^2+(\varepsilon+\lambda_+)^2)$ of a hidden-driven mode whose own variance stays of order one.

\section{High-dimensional self-averaging and edge sensitivity}

So far, we have focused on the role of focal--hidden couplings and, consequently, on how the hidden-sector self-energy correction modifies the observable traits. For a stabilizing hidden sector, we found that diversification of the observable traits occurs when the self-energy correction is sufficiently strong to overcome the stabilizing margin of the focal sector. We now ask how this correction behaves as the dimensionality of the hidden trait space increases. In particular, does the self-energy approach a finite limit, or does it grow or decrease without bound as the number of hidden dimensions increases? For a single focal trait and an $h$-dimensional hidden sector, Eq.~\eqref{eq:modes} gives
\begin{equation}
\Sigma=-\sum_{\alpha=1}^{h}\frac{b_\alpha^2}{\mu_\alpha}.
\label{eq:sigmah}
\end{equation}
For a stabilizing hidden sector, $\mu_\alpha<0$ for all $\alpha\in\{1,2,\dots,h\}$, so every hidden mode contributes positively to $\Sigma$. If each of the $h$ hidden modes has an order-one coupling and the hidden eigenvalues remain of order one, the total correction generally grows linearly with $h$. A nontrivial high-dimensional limit therefore requires the coupling strength per mode to decrease with dimension. In particular, choosing $b_\alpha\asymp h^{-1/2}$ gives $b_\alpha^2\asymp h^{-1}$, so that the cumulative contribution of $h$ modes can remain of order one, provided the $\mu_\alpha$ remain of order one.

To obtain a controlled large-$h$ limit, consider a stabilized hidden block
\begin{equation}
\Mww=-\delta I_h+W_h,
\label{eq:Mww}
\end{equation}
where $W_h$ is a centered symmetric random matrix whose eigenvalues follow a semicircle of radius $R$, and let the focal--hidden coupling vector have entry variance $R^2/(4h)$. This choice makes all off-diagonal entries of $M$, focal--hidden and hidden--hidden alike, statistically identical, with variance $R^2/(4h)$: the focal trait is distinguished only by being measured.
For $\delta>R$,
\begin{equation}
\begin{aligned}
\Sigma_h&=\bm v\T(\delta I_h-W_h)^{-1}\bm v\;\xrightarrow[h\to\infty]{}\;\Sigma_\infty(\delta),\\
\Sigma_\infty(\delta)&=\frac{R^2}{4}\,g_{\rm sc}(\delta)=\tfrac12\Big(\delta-\sqrt{\delta^2-R^2}\Big),
\end{aligned}
\label{eq:sigmainf}
\end{equation}
(see Appendix~\ref{app:selfenergy}) with the semicircle Stieltjes transform
\begin{equation}
g_{\rm sc}(\delta)=\frac{2}{R^2}\Big(\delta-\sqrt{\delta^2-R^2}\Big).
\label{eq:gsc}
\end{equation}
Thus, for a single focal trait with $\Muu=-\varepsilon$, diversification occurs when the hidden-sector correction exceeds the focal stabilizing strength, $\Sigma_\infty>\varepsilon$. In the high-dimensional limit, the transition becomes sharp: away from the threshold, diversification either occurs with probability close to one or is almost absent. Increasing the number of hidden traits sharpens this transition but does not by itself guarantee diversification. This is not in tension with the finding of \citet{doebeli2010} that diversification becomes almost certain in high dimensions: with order-one couplings, as there, the upper spectral edge of $M$ grows as $2\sigma\sqrt h-\varepsilon$ and eventually crosses zero for any $\varepsilon$, whereas dimension-normalized couplings give a finite limit. The two statements describe different scalings of coupling strength with trait number, and which applies to a given organism is an empirical question.

\begin{figure*}[t]
\centering
\includegraphics[width=\textwidth]{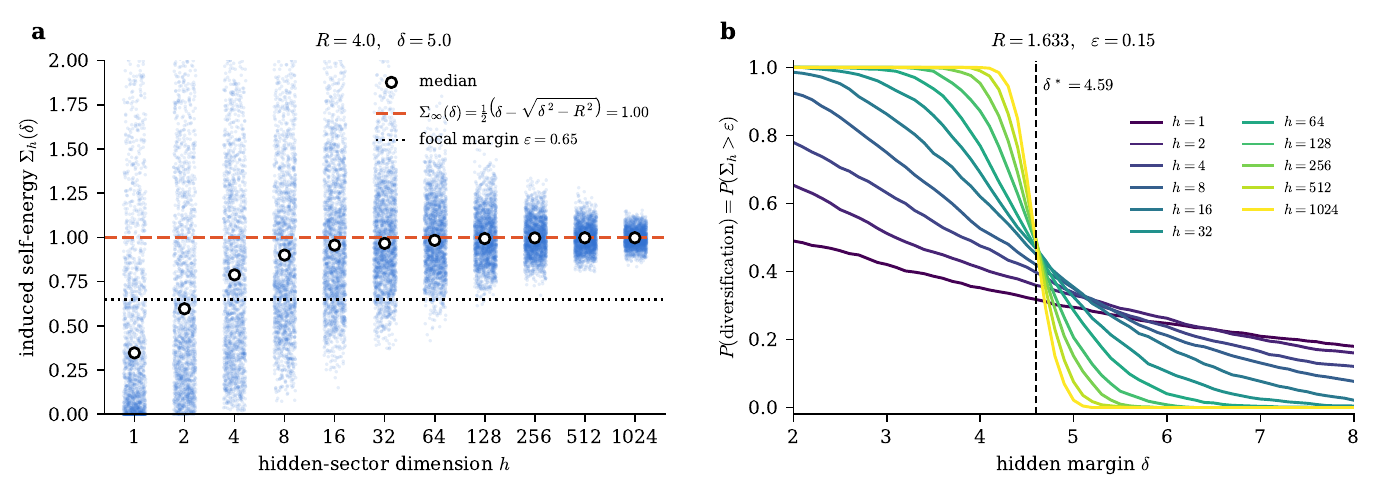}
\caption{\textbf{Self-energy and sharpening of the diversification threshold with increasing hidden-sector dimension:} (a) As the dimension of the hidden sector increases, the self-energy correction to the focal trait converges to a deterministic value, with fluctuations across random-matrix realizations decreasing with dimension. For each dimension, 2000 random-matrix realizations are shown as scatter points, with the median indicated by a white circle. The median approaches the theoretically predicted value as the dimension increases. The parameter values are $R=4.0$, $\varepsilon=0.65$, and $\delta=5.0$. (b) In a high-dimensional random hidden sector, dimension-normalized couplings produce a finite self-energy limit given by Eq.~\eqref{eq:sigmainf}. Diversification of the focal trait occurs when this correction exceeds the focal stabilizing margin. As the hidden-sector dimension increases, the diversification probability changes more sharply from one to zero as the hidden margin crosses $\delta^\ast$, resulting in an increasingly sharp onset of diversification. The parameter values are $R=1.6330$ and $\varepsilon=0.15$.}
\label{fig:selfenergy}
\end{figure*}

It is evident from Eq.~\eqref{eq:gsc} that the self-energy is a decreasing function of the hidden margin. Diversification requires $\Sigma_\infty>\varepsilon$, so there is a critical hidden margin

\begin{equation}
\delta^\ast=\varepsilon+\frac{R^2}{4\varepsilon},
\qquad \varepsilon\le R/2,
\label{eq:deltastar}
\end{equation}
beyond which the hidden sector cannot catalyze diversification however large $h$ becomes, and below which it does so with probability approaching one (Fig.~\ref{fig:selfenergy}b). For $\varepsilon>R/2$ the self-energy never reaches $\varepsilon$, because its largest value, attained at the spectral edge $\delta=R$, is $R/2$; the hidden sector then cannot catalyze diversification at any margin.

Thus far, we have considered Gaussian random-matrix ensembles for the competition and stabilizing operators. It is useful to place this analysis in the broader context of established random-matrix spectral transitions. If $A$ and $K$ are modeled as Wishart matrices, the generalized eigenvalues of the pair $(A,K)$ are related to the eigenvalues of a Fisher/Jacobi-type random-matrix ensemble. The largest generalized eigenvalue is then the analogue of Roy's largest root, whose edge fluctuations can exhibit Tracy--Widom statistics under appropriate asymptotic conditions \citep{roy1953,tracy1994,johnstone2008}.

A fixed-rank focal structure or structured focal--hidden coupling can similarly be viewed as a low-rank deformation of the random background. In spiked random-matrix ensembles, such perturbations can produce outlier eigenvalues and eigenvector alignment through BBP-type transitions \citep{baik2005,benaych2011,johnstone2017,wang2017}. This provides a useful spectral analogy for the emergence of a distinguished diversifying direction in the present model.

The operator $K-A=-M$ may also be viewed as a random Hessian-like operator for the local stability of the reference phenotype. The loss of local stability occurs when its lowest eigenvalue crosses zero, which connects the present spectral criterion to the broader theory of topology changes and complexity transitions in random landscapes \citep{fyodorov2004,fyodorov2012,auffinger2013,ros2019}. Finally, the May--Wigner transition in ecological Jacobians provides a related spectral transition in ecological stability, although the correspondence is not exact because ecological Jacobians are generally non-symmetric whereas the operator $K-A$ considered here is symmetric \citep{may1972,allesina2012,fyodorov2016}.

\section{Sequential elimination as an exact decimation}

Analogous to Wilsonian decimation, integrating out one trait renormalizes the couplings among the remaining traits, with the margin $\delta=-M_{rr}>0$ controlling the strength of the induced correction. Specifically, eliminating a single trait $r$ transforms every remaining matrix element by
\begin{equation}
M_{ij}\longmapsto M_{ij}-\frac{M_{ir}M_{rj}}{M_{rr}}.
\label{eq:decimate}
\end{equation}
Repeated elimination of traits is equivalent to taking the Schur complement with respect to the entire eliminated block. Even though the intermediate couplings in general depend on the order of elimination of traits, the final reduced operator is independent of the choice of order. A natural choice for decimation is to rank the stable modes by their margins $\delta=-M_{rr}>0$ and eliminate them from the largest to the smallest. In this scheme, the near-marginal modes are retained until late in the flow.

Let $\delta_c$ be the cutoff, such that all stable modes with $\delta>\delta_c$ are eliminated and all modes with $\delta\le\delta_c$ are retained. If $\rho(\delta)$ is the density of the stable modes per unit margin and the coupling variance is independent of the margin, with value $R^2/4$, then the cumulative correction due to the eliminated modes is
\begin{equation}
\Sigma(\delta_c)=\frac{R^2}{4}\int_{\delta_c}^{\infty}\frac{\rho(\delta)}{\delta}\,d\delta .
\label{eq:flow}
\end{equation}
With logarithmic scale, $\ell=-\ln\delta_c$,
\begin{equation}
\frac{d\Sigma}{d\ell}=\frac{R^2}{4}\,\rho(\delta_c).
\label{eq:flowode}
\end{equation}
This flow equation clarifies the behaviour near the marginal edge. Although the $1/\delta$ kernel in Eq.~\eqref{eq:flow} makes near-marginal modes disproportionately important, their aggregate contribution does not diverge at the collective edge in the present case. In the large-$h$ limit, the semicircular spectral density vanishes as a square root at the edge, $\sqrt{\delta-R}$. Consequently the integrand in Eq.~\eqref{eq:flow} remains integrable at the edge. This is further emphasized by Eq.~\eqref{eq:flowode}: the rate of change of the self-energy correction is directly proportional to the mode density, and as the spectral edge is approached $\rho(\delta_c)\to0$, hence $d\Sigma/d\ell\to0$. Therefore the correction to the self-energy due to further elimination slows down near the edge, causing the self-energy to reach a finite limiting value rather than diverge.

The collective spectral edge requires a distinction from the pole associated with an individual mode. In the large-$h$ limit, the self-energy $\Sigma_\infty$ remains finite as $\delta\downarrow R$, but its derivative diverges with a square-root singularity. For finite systems, however, individual realizations can exhibit resonant peaks when a hidden eigenvalue lies unusually close to zero. Accordingly, the robust large-system prediction is strong edge sensitivity and enhanced finite-size fluctuations, not a divergent ensemble mean, as shown in Fig.~\ref{fig:selfenergy}a.

\section{Conclusion and Future Directions}

In this work, we are broadly interested in the role of unobservable traits in the evolution of observable traits. Several phenotypic traits may be interconnected through ecological interactions, yet it is not obvious how such complex interactions influence the process of diversification. Our findings show that partial observation can fundamentally alter how adaptive diversification is inferred from phenotypic data. By partitioning the phenotype space into focal and hidden sectors and eliminating the hidden sector, we obtain a modified effective operator in the focal space whose correction is determined by the hidden-sector resolvent. This self-energy provides a unified description of three effects that would otherwise appear unrelated: stable hidden traits can catalyze diversification in focal traits that are stable in isolation, diversifying hidden traits drive diversification that the observable sector displays but cannot account for, and hidden-mediated interactions can generate correlations between focal traits that have no direct coupling. The direction of these effects is determined by the coupling between the focal and hidden traits, while their magnitude becomes particularly sensitive as a hidden mode approaches marginality.

The analysis also shows that the number of diversifying directions is a property of the full competition-minus-stabilization operator rather than of the measured traits alone. In particular, a diversification transition corresponds to a crossing of the upper spectral edge of $M=A-K$ through zero, while block elimination determines how this transition is represented in a restricted observable subspace. In high-dimensional hidden sectors, dimension-normalized couplings lead to a finite self-energy and to a sharpening of the diversification transition through self-averaging. Thus, increasing the number of unobserved traits does not simply increase unexplained variability; it can systematically renormalize the effective evolutionary geometry seen in the measured traits.

Individual-based simulations support the principal predictions of the theory. In the one-dimensional Gaussian model, the simulations recover both the diversification threshold and the predicted maintained-variance scale. Hidden-trait coupling increases focal variation in the predicted direction, while the quartic model provides a more direct signature through coupling-induced trait splitting. Moreover, reversing the sign of one hidden--focal coupling reverses the sign of the induced correlation between otherwise uncoupled focal traits. These results support the interpretation that hidden interactions determine not only whether diversification occurs, but also whether the diversification displayed by a restricted set of measured traits can be attributed to those traits at all.

\paragraph{Biological consequences}
Several predictions follow that are not visible in a treatment that regards all traits as observed. \emph{First}, whether a measured trait diversifies is not a property of that trait's own ecology. Measuring $a_{uu}<k_{uu}$ and concluding that stabilizing selection precludes diversification is unsafe, because the hidden correction can take either sign and is unbounded near a hidden margin; conversely, diversification in a trait whose measured ecology says it should be stable is the expected signature of a hidden driver, not evidence of mismeasurement. \emph{Second}, correlations between measured traits in a diversifying population need not reflect any direct interaction between them: a shared hidden trait induces a correlation of sign $c_1c_2$, and reversing one coupling reverses it. This is a latent-variable prediction with an ecological, frequency-dependent origin rather than a pleiotropic one, and it is testable by manipulating the hidden trait. \emph{Third}, the traits most likely to be driving an observed diversification are those under the weakest net selection. Near-marginal hidden traits are the lever arms of Eq.~\eqref{eq:modes}, with contribution $\propto1/|\mu_\alpha|$; a scan that ranks candidate traits by the strength of selection on them will rank these last. \emph{Fourth}, a hidden driver cannot be ruled out from focal data alone, but it can be tested by manipulation: stabilizing the focal trait extinguishes catalyzed diversification at a finite threshold and only attenuates hidden-driven diversification. By contrast, measuring a candidate hidden trait and finding an association with the focal traits is consistent with either regime. \emph{Fifth}, in organisms with many weakly coupled hidden traits the outcome becomes predictable rather than idiosyncratic: the correction self-averages to a deterministic value set by two numbers, the aggregate coupling variance and the typical hidden margin, and diversification becomes a threshold phenomenon in the latter (Eq.~\eqref{eq:deltastar}). \emph{Sixth}, the decimation ordering gives a principled answer to which traits an empiricist should measure given a limited budget: those closest to their own thresholds, not those with the largest effects. \emph{Finally}, when the unmeasured sector is the unstable one, no model confined to the measured traits can predict their diversification; the effective theory lives in the hidden coordinates and the measured traits are passengers. This is a limit on inference, not on observation.

Although the focal--hidden reduction itself is exact, the present derivation is formulated for the quadratic competition-minus-stabilization operator. A natural direction for future work is therefore to extend the exact hidden-trait elimination beyond the quadratic description of the competition and stabilizing landscapes. In particular, higher-order terms could be incorporated to determine how the hidden-sector self-energy is modified away from the quadratic regime and how the effective focal description evolves beyond the diversification threshold. Such an extension would provide a direct analytical connection between the present Gaussian theory and the cluster-forming behavior observed in the quartic model, to which the quadratic criterion does not directly apply, while retaining the central idea of eliminating unobserved phenotypic degrees of freedom.

A second direction is to investigate stochastic and finite-population effects systematically. The present theory predicts the location and orientation of the diversification transition, whereas the amplitude of the resulting variation depends on demographic noise and nonlinear ecological feedback. A finite-population theory could therefore determine how the deterministic spectral transition is rounded at finite population size and how the observed variance and cluster number scale with population size.

Finally, the theory suggests a concrete empirical strategy: rather than treating unmeasured traits simply as sources of statistical uncertainty, one can ask whether latent trait combinations close to their own diversification threshold predict enhanced variation, induced correlations, or apparent diversification in measured traits. Experimental systems in which multiple phenotypic dimensions can be measured simultaneously could therefore be used to test whether the focal variation is a correlated response to a hidden trait. Conditioning on candidate hidden traits identifies the association but does not by itself separate catalysis from masking, since in both the stationary population is concentrated along a single direction; the discriminating manipulation is to stabilize the focal trait, which extinguishes catalyzed diversification at a finite threshold but only attenuates hidden-driven diversification. More generally, the framework provides a route toward distinguishing diversification that is intrinsic to an observed trait from diversification that is generated by, or driven from, phenotypic dimensions that have not been measured.

Taken together, the results establish hidden-trait elimination as a general mechanism by which incomplete phenotypic observation can reshape the apparent evolutionary dynamics. The central object is not the focal trait alone, but the effective operator obtained after accounting for the hidden sector. This perspective provides a bridge between multidimensional evolutionary dynamics, effective theories of partially observed systems, and spectral descriptions of high-dimensional diversification.

An important direction for future experimental work is to test whether coupling to an unobserved trait can induce diversification in an otherwise stable observable trait. Although, to our knowledge, such a direct experimental test is currently lacking, several studies point to closely related phenomena in which latent biological variation can influence observable phenotypic diversification. In \emph{Myxococcus xanthus}, experimental evolution demonstrated that phenotypic diversification can remain latent under one environmental condition and become expressed under another \citep{rendueles2020}. Similarly, experiments in tomato showed that combinations of cryptic genetic variants can interact to produce substantial changes in inflorescence architecture \citep{zebell2025}. While these studies do not directly test the hidden--focal coupling mechanism developed here, they highlight the broader experimental relevance of latent variation as a source of observable phenotypic change. A direct experimental test of the mechanism proposed in this work would therefore provide an important next step toward connecting hidden-trait effects with observable patterns of adaptive diversification.

% ============================================================
\section*{Analytical details}
\appendix
\renewcommand{\thesubsection}{A\arabic{subsection}}
\titleformat{\subsection}{\normalfont\sffamily\bfseries}{\thesubsection}{0.6em}{}
\setcounter{section}{0}

\subsection{Gaussian equilibrium and the maintained-variance law}
\label{app:gauss}

Assume a stationary density $\phi^\ast(\bm x)\propto\exp(-\bm x\T B\bm x)$. Gaussian convolution gives
\begin{equation}
\int\! a(\bm x,\bm y)\phi^\ast(\bm y)\,d\bm y\propto\exp\!\big\{-\bm x\T[A-A(A+B)^{-1}A]\bm x\big\}.
\end{equation}
Matching this exponent to the carrying capacity yields
\begin{align}
B&=A(A-K)^{-1}A-A\\
&=K+K(A-K)^{-1}K .
\label{eq:B}
\end{align}
When $A-K\succ0$, $B\succ0$ and the covariance is $(2B)^{-1}=\tfrac12(K^{-1}-A^{-1})$. In one dimension, this reduces to Eq.~\eqref{eq:var1d}.

When $A-K$ has mixed inertia, $\exp(-\bm x\T B\bm x)$ is not normalizable and Eq.~\eqref{eq:B} does not define a covariance. The stationary state follows instead from the decoupling of \citet{doebeli2010}: let $S$ satisfy $S\T AS=I$ and $S\T KS=D=\operatorname{diag}(\kappa_1,\dots,\kappa_m)$. In the coordinates $\bm y=S^{-1}\bm x$ the dynamics separates into independent one-dimensional models with $a=1$ and $k=\kappa_i$, so each direction with $\kappa_i<1$ carries variance $(1-\kappa_i)/(2\kappa_i)$ and each direction with $\kappa_i\ge1$ collapses to the optimum, broadened only by mutation. Hence
\begin{equation}
\operatorname{Cov}(\bm x)=\tfrac12\,S\,[D^{-1}-I]_+\,S\T,
\label{eq:covproj}
\end{equation}
where $[\cdot]_+$ keeps only the positive diagonal entries. For $A-K\succ0$ every entry is positive and Eq.~\eqref{eq:covproj} reduces to $(2B)^{-1}$; in general the stationary measure is Gaussian on the diversifying subspace and concentrated along the stabilized directions.

Consider a system consisting of one focal trait $x$ coupled to one hidden trait $h$. For simplicity, let the competition matrix be $A=I$, and write
\begin{equation}
M=A-K=\begin{pmatrix}-\varepsilon&c\\c&-\delta\end{pmatrix},\qquad\varepsilon>0,
\label{eq:toy2}
\end{equation}
where $\varepsilon$ characterizes the subthreshold curvature of the focal trait, $\delta$ characterizes the stability of the hidden trait ($\delta>0$ stable, $\delta<0$ diversifying on its own), and $c$ denotes the coupling between the focal and hidden traits. Since $K=A-M$, the carrying-capacity matrix is
\begin{equation}
K=\begin{pmatrix}1+\varepsilon&-c\\-c&1+\delta\end{pmatrix}.
\end{equation}
With $A=I$, $S$ is the orthogonal eigenbasis of $K$ and $\kappa_i=1-\lambda_i(M)$, so a direction diversifies iff the corresponding eigenvalue of $M$ is positive. Whenever $\det M=\varepsilon\delta-c^2<0$, $M$ has exactly one positive eigenvalue,
\begin{equation}
\lambda_+=\tfrac12\Big[-(\varepsilon+\delta)+\sqrt{(\varepsilon-\delta)^2+4c^2}\,\Big],
\end{equation}
with unit eigenvector $\bm v\propto(c,\ \varepsilon+\lambda_+)\T$. The stationary population is Gaussian along $\bm v$ with variance $\tfrac12\lambda_+/(1-\lambda_+)$ and concentrated across it, and the focal-trait variance is the $(1,1)$ entry of Eq.~\eqref{eq:covproj},
\begin{equation}
\operatorname{Var}(x)=\frac12\,\frac{\lambda_+}{1-\lambda_+}\,\frac{c^2}{c^2+(\varepsilon+\lambda_+)^2},
\end{equation}
which is Eq.~\eqref{eq:var2d}. For $\delta>0$ it vanishes continuously at the threshold $c^2=\varepsilon\delta$; for $\delta<0$ it is positive for every $c\neq0$ and every $\varepsilon$, which is the masking regime. (The $(1,1)$ entry of $\tfrac12(K^{-1}-I)$ differs from this by the contribution of the stabilized direction and is not a variance when $M$ is indefinite.)

\subsection{Diversification at the pole}
\label{app:pole}

 When the hidden block $\Mww$ is singular, the ordinary Schur complement is not defined. 

Let the complete operator be
\begin{equation}
M(\mu)=\begin{pmatrix}\Muu&\Muw\\ \Mwu&\Mww(\mu)\end{pmatrix},
\end{equation}
where $\Mww(\mu)$ has a simple eigenvalue $\mu$ with normalized eigenvector $\bm v$. Assume that the remaining hidden eigenvalues stay bounded away from zero. In an orthonormal hidden basis beginning with $\bm v$, write
\begin{equation}
\Mww(\mu)=\begin{pmatrix}\mu&0\\0&M_{ww}^{\perp}\end{pmatrix},\qquad
\Muw=\begin{pmatrix}\bm b&M_{uw}^{\perp}\end{pmatrix},
\end{equation}
where
\begin{equation}
\bm b=\Muw\bm v\in\mathbb R^p .
\end{equation}
The noncritical hidden modes may be eliminated first. Define
\begin{equation}
\widehat M_{uu}=\Muu-M_{uw}^{\perp}(M_{ww}^{\perp})^{-1}M_{wu}^{\perp}.
\end{equation}
The quotient property of Schur complements gives an exact reduced operator on the focal sector plus the single critical hidden coordinate:
\begin{equation}
\widehat M(\mu)=\begin{pmatrix}\widehat M_{uu}&\bm b\\ \bm b\T&\mu\end{pmatrix}.
\end{equation}
For $\mu\neq0$, eliminating the final hidden coordinate gives
\begin{equation}
\Meff(\mu)=\widehat M_{uu}-\frac{\bm b\bm b\T}{\mu}.
\end{equation}
All singular behavior associated with the simple hidden threshold is therefore contained in one explicit rank-one term. The regular hidden sector changes the background focal operator from $\Muu$ to $\widehat M_{uu}$, but it does not alter the pole structure.

This reduction also clarifies the meaning of near-criticality. A hidden sector may be high-dimensional, yet only the hidden eigenvectors that both approach zero and couple through $\Muw$ contribute singularly. Dimension alone does not generate the pole.

If $\widehat M_{uu}$ is invertible, the inertia exactly at the hidden threshold follows by eliminating the focal block rather than the singular hidden block:
\begin{equation}
\In\,\widehat M(0)=\In(\widehat M_{uu})+\In\!\left(-\bm b\T\widehat M_{uu}^{-1}\bm b\right).
\label{eq:poleinertia}
\end{equation}
For $\widehat M_{uu}\prec0$ and $\bm b\neq0$, the scalar in Eq.~\eqref{eq:poleinertia} is positive, so the complete system has one positive direction at $\mu=0$. For $\widehat M_{uu}\succ0$, it is negative, so the marginal mode contributes a stabilized direction and the number of positive directions is unchanged. The full operator is generically indefinite rather than singular when the hidden mode alone reaches zero.

\subsection{Number of diversifying directions and hidden traits}
\label{app:count}

Let us first consider one stabilizing hidden mediator, i.e., $\Mww=-\Delta$ where $\Delta>0$ is a constant. Let the focal sector $\Muu=-P$, $P\succ0$, be of arbitrary dimension $p\times p$ and stabilizing. Therefore after eliminating the hidden trait the effective focal operator is
\begin{equation}
\Meff=-P+\frac{\Muw\Muw\T}{\Delta}.
\end{equation}
Define the stiffness-normalized coupling as
\begin{equation}
\bm\alpha=\frac{P^{-1/2}\Muw}{\sqrt\Delta},
\end{equation}
then
\begin{equation}
\Meff=P^{1/2}\left(-I_p+\bm\alpha\bm\alpha\T\right)P^{1/2},
\end{equation}
where $I_p$ is the $p$-dimensional identity matrix. Since congruence preserves inertia, the numbers of positive and negative eigenmodes are the same for $(-I_p+\bm\alpha\bm\alpha\T)$ and $\Meff$. The normalized matrix $-I_p+\bm\alpha\bm\alpha\T$ has eigenvalue $-1+\|\bm\alpha\|^2$ along $\bm\alpha$ and eigenvalue $-1$ on its orthogonal complement. Hence it has one positive eigenvalue precisely when
\begin{equation}
\|\bm\alpha\|^2=\frac{\bm c\T P^{-1}\bm c}{\Delta}>1 .
\end{equation}
Now, suppose that the hidden block is stabilizing and of arbitrary dimension:
\begin{equation}
\Mww=-H,\qquad H\succ0,
\end{equation}
then the normalized stiffness-coupling matrix is
\begin{equation}
C=P^{-1/2}\Muw H^{-1/2},
\end{equation}
and the effective focal matrix
\begin{equation}
\Meff=P^{1/2}\left(-I_p+CC\T\right)P^{1/2}.
\end{equation}
Let $\sigma_1(C)\ge\sigma_2(C)\ge\cdots$ denote the singular values of $C$. The number of interaction-induced positive focal directions is
\begin{equation}
n_+(\Meff)=\#\{a:\sigma_a(C)>1\}.
\end{equation}
In particular, a stable hidden sector of effective rank $r$ can induce at most $r$ independent focal diversifying directions in a focal sector that is stable on its own. (Without the hypothesis $\Muu\prec0$ the general bound is $n_+(\Meff)\le n_+(\Muu)+\operatorname{rank}\Muw$.) Therefore, one hidden trait produces one normalized coupling vector and can create at most one positive channel. Several independent hidden coupling profiles can create several focal channels, but only when their stiffness-normalized singular values cross unity.

\subsection{High-dimensional self-energy for dimension-normalized coupling}
\label{app:selfenergy}

Let us consider a random coupling matrix
\begin{equation}
\Muw=\frac1{\sqrt h}\,S^{1/2}G_h,
\label{eq:Gh}
\end{equation}
where $h$ is the dimension of the hidden sector. Here, $G_h\in\mathbb R^{p\times h}$ has independent standard Gaussian entries and is independent of $\Mww$. The symmetric positive definite matrix $S$ encodes the total focal coupling covariance. Substituting Eq.~\eqref{eq:Gh} into the self-energy gives
\begin{equation}
\Sigma_h=\frac1h\,S^{1/2}G_h\,\mathcal R_h(\delta)\,G_h\T S^{1/2},
\end{equation}
where $\mathcal R_h(\delta)$ is the resolvent of $\Mww$ defined in Eq.~\eqref{eq:Mww}. Conditioned on $\Mww$, the resolvent $\mathcal R_h(\delta)$ is fixed and independent of $G_h$. Therefore,
\begin{equation}
\mathbb E[\Sigma_h\,|\,\Mww]=\frac1h\,S^{1/2}\,\mathbb E\!\left[G_h\mathcal R_h(\delta)G_h\T\right]S^{1/2}.
\end{equation}
Using the Gaussian second-moment identity $\mathbb E[G_{ia}G_{jb}]=\delta_{ij}\delta_{ab}$, we obtain
\begin{equation}
\begin{aligned}
\mathbb E\!\left[(G_h\mathcal R_hG_h\T)_{ij}\right]&=\sum_{a,b}(\mathcal R_h)_{ab}\,\mathbb E[G_{ia}G_{jb}]\\
&=\delta_{ij}\sum_a(\mathcal R_h)_{aa}=\delta_{ij}\operatorname{tr}\mathcal R_h .
\end{aligned}
\end{equation}
Thus,
\begin{equation}
\mathbb E\!\left[G_h\mathcal R_hG_h\T\right]=\operatorname{tr}(\mathcal R_h)\,I_p,
\end{equation}
and hence
\begin{equation}
\mathbb E[\Sigma_h\,|\,\Mww]=\frac{\operatorname{tr}\mathcal R_h(\delta)}{h}\,S=m_h(\delta)\,S,
\end{equation}
where $m_h(\delta)=h^{-1}\operatorname{tr}\mathcal R_h(\delta)$. We can therefore write the self-energy as
\begin{equation}
\Sigma_h=m_h(\delta)S+\left[\Sigma_h-m_h(\delta)S\right],
\end{equation}
where the first term is the conditional mean and the second represents the fluctuations due to the random focal--hidden couplings. Under the spectral convergence, we have
\begin{equation}
m_h(\delta)\longrightarrow g_\nu(\delta)=\int\frac{1}{\delta-x}\,\nu(dx).
\end{equation}
Moreover, with the $h^{-1/2}$ normalization in Eq.~\eqref{eq:Gh}, the self-energy is a sum of $h$ independent hidden-mode contributions, each of order $h^{-1}$. Consequently, their cumulative fluctuations decrease as $h^{-1/2}$, provided the resolvent remains uniformly bounded by the spectral gap. Thus the fluctuation term vanishes in mean square as $h\to\infty$, and
\begin{equation}
\Sigma_h(\delta)\longrightarrow\Sigma_\infty(\delta)=g_\nu(\delta)\,S,
\end{equation}
which establishes the self-averaging of the hidden-sector correction.

\subsection{Diversification in three focal and three hidden traits}
\label{app:three}

\begin{figure*}[t]
\centering
\includegraphics[width=\textwidth]{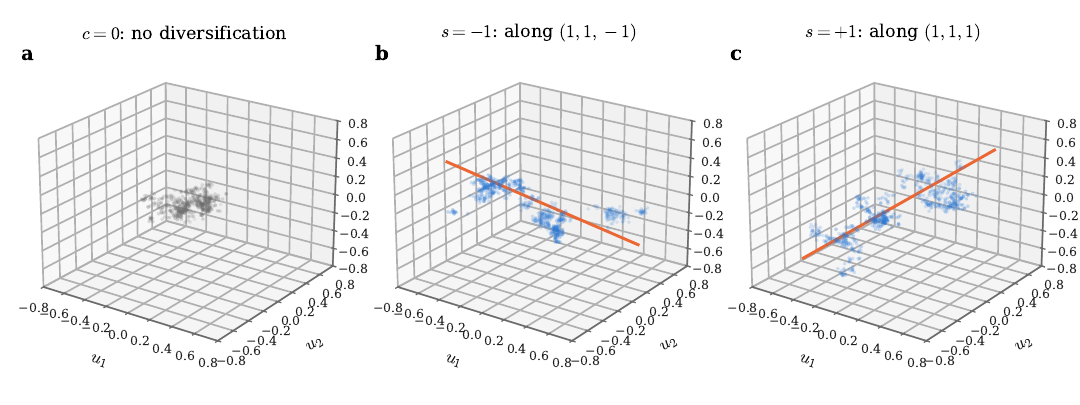}
\caption{\textbf{Diversification in three focal and three hidden traits:} Subplot (a) corresponds to the uncoupled case ($c=0$), where diversification does not occur because both the focal and hidden traits are individually stable. Subplot (b) shows diversification along the $(1,1,-1)$ direction for $s=-1$, whereas subplot (c) shows diversification along the $(1,1,1)$ direction for $s=1$. The plots are obtained after $3\times10^6$ birth--death events with $\varepsilon=0.1$, $\delta=0.2$, and $c=0.2$, in the presence of mutation ($\sigma_{\rm mut}=0.015$); orange lines mark the predicted directions, and the leading eigenvector of the measured focal covariance is within $3^\circ$ of each.}
\label{fig:three}
\end{figure*}

We consider three focal traits $\bm u\in\mathbb R^3$ coupled to three hidden traits $\bm w\in\mathbb R^3$. Let
\begin{equation}
M=\begin{pmatrix}-\varepsilon I_3&C\\C\T&-\delta I_3\end{pmatrix},\qquad
C=c\begin{pmatrix}1&1&1\\1&1&1\\s&s&s\end{pmatrix},
\end{equation}
where $s=\pm1$ is a free sign. The hidden-sector elimination gives the exact effective focal operator
\begin{equation}
\Meff=-\varepsilon I_3+\frac1\delta CC\T .
\end{equation}
Since
\begin{equation}
CC\T=3c^2\begin{pmatrix}1&1&s\\1&1&s\\s&s&1\end{pmatrix},
\end{equation}
the effective operator becomes
\begin{equation}
\Meff=-\varepsilon I_3+\frac{3c^2}{\delta}\begin{pmatrix}1&1&s\\1&1&s\\s&s&1\end{pmatrix}.
\end{equation}
Its largest eigenvalue is
\begin{equation}
\lambda_{\max}(\Meff)=-\varepsilon+\frac{9c^2}{\delta},
\end{equation}
so diversification occurs when
\begin{equation}
c^2>\frac{\varepsilon\delta}{9}.
\end{equation}
The corresponding focal diversification direction is
\begin{equation}
\bm v_\parallel=\frac1{\sqrt3}(1,1,s)\T,
\end{equation}
which gives $(1,1,1)$ for $s=+1$ and $(1,1,-1)$ for $s=-1$. The two orthogonal focal modes remain stable with eigenvalue $-\varepsilon$ (see Fig.~\ref{fig:three}).

% ============================================================

\medskip\small\noindent We acknowledge financial support from the Anusandhan National Research Foundation (ANRF) under Grant No. ANRF/ARGM/2025/002679/TS.

We also acknowledge the use of AI-based tools to assist in improving the clarity and presentation of the manuscript, refining schematic figures, and cross-checking the simulation code and numerical results. All scientific ideas, analyses, interpretations, and conclusions presented in this work are the sole responsibility of the authors.

\end{document}